\documentclass[
    ,final            
  ]
  {aipproc}

\layoutstyle{8x11single}

\begin{document}

\title{Binaries and GLAST}


\author{Guillaume Dubus}{
  address={Laboratoire d'Astrophysique de Grenoble, Universit\'e J. Fourier, CNRS, BP 53, 38041 Grenoble, France}
}

\begin{abstract}
  Radio and X-ray observations of the relativistic jets of
  microquasars show evidence for the acceleration of particles to very
  high energies. Signatures of non-thermal processes occurring closer
  in to the compact object can also be found. In addition, three
  binaries are now established emitters of high ($>$100~MeV) and/or
  very high ($>$100~GeV) energy gamma-rays. High-energy emission can
  originate from a microquasar jet (accretion-powered) or from a
  shocked pulsar wind (rotation-powered). I discuss the impact GLAST
  will have in the very near future on studies of such binaries. GLAST
  is expected to shed new light on the link between accretion and
  ejection in microquasars and to enable to probe pulsar winds on
  small scales in rotation-powered binaries.
\end{abstract}

\maketitle


\section{Introduction}
Binaries composed of a black hole or neutron star in orbit with a
stellar companion are prominent sources of the X-ray sky.  X-ray
binaries are usually powered by accretion of matter from the
companion. The gravitational energy released by accretion
heats the plasma to temperatures in the range 1~keV~$ < kT <$ 50~MeV
for a stellar-mass black hole, which is someway below the GLAST energy
range.  However, part of the power is also emitted non-thermally by
particles of much greater energies. The radio emission from X-ray
binaries, due to synchrotron radiation from electrons located in
relativistic outflows, provides clear evidence for this.

These relativistic jets are the most striking demonstration of the
analogy between accretion onto the stellar-mass compact objects in
X-ray binaries and onto the supermassive black holes in Active
Galactic Nuclei (AGN). X-ray binaries with relativistic jets have thus
been denominated `microquasars'. Similarities also exist in timing and
spectral characteristics. The conjecture, at least for black holes, is
that the underlying scaling factor of the physical processes is the
gravitational radius $R_g$=$GM/c^2$ (and its associated timescale
$R_g/c$).

The similarities have prompted speculation that some X-ray binaries
may be analogs of blazars, AGNs dominated by non-thermal output
because their relativistic jet is fortuitously aligned with the
line-of-sight. Blazars emit a large fraction of their power in high
energy (HE) gamma-rays ($>$100 MeV). Notwithstanding the possible
issue that the microquasar population may be too small for a chance
alignment to occur, finding genuine `microblazars' could also prove
exceedingly difficult if the scaling strictly holds, since the
observed sub-hour TeV flaring in blazars translates to milliseconds
for a microblazar. But even if the jet is misaligned, the high
particle energies implied by the radio to X-ray observations make it
likely, if not unavoidable, that HE gamma-ray emission should be
present at some level in generic microquasars.  Cataclysmic variables
(AE Aqr) and colliding winds in Wolf-Rayet binaries have also been
proposed to emit HE gamma-rays - but these systems will not be
addressed here.

GLAST observations will soon open a new window into non-thermal
processes in accreting binaries. Three compact binaries are presently
known sources of HE gamma-rays. These three {\em gamma-ray binaries}
(in that most of their radiative output is in gamma-rays, regardless
of the underlying physics) are probably all rotation-powered by the
spin-down of a young pulsar rather than accretion-powered.  The current
observational status and GLAST prospects for both types of sources are
reviewed.

\section{GLAST studies of accreting binaries}

\subsection{X-ray jets} The best evidence for particle acceleration to
high energies in accreting binaries comes from the observation of
X-ray emission from localized regions in the relativistic jet of two
microquasars, XTE J1550-563 and H1743-322
\cite{2002Sci...298..196C}. The radio spectrum
from these regions connects remarkably well with the X-ray spectrum
across nine decades in energy. The overall spectral slope
$\alpha\approx -0.6$ points to synchrotron emission from a canonical
non-thermal distribution of electrons $dN\propto N^{-2.2}dE$ over nine
orders-of-magnitude. The emission zone is resolved ($\approx$1''). The
equipartition magnetic field is $\approx 500~\mu$G and the X-ray
emitting electrons have energies $\approx 10$~TeV
\cite{2002Sci...298..196C}.

Electrons of such energies necessarily emit in the GLAST energy range.
However, the prospects for detecting this emission are not favorable.
Synchrotron losses limit the maximum possible electron energy to a few
PeV, in which case the spectrum could extend to a 100~MeV at a level
$\approx 10^{-12}$~ergs~s$^{-1}$~cm$^{-2}$, too faint for GLAST
observations in the Galactic Plane.  Moreover, the short radiative
timescale ($\approx 5$~days) of PeV electrons would lead to a break or
cutoff in the spectrum below 100~MeV. (The timescale also shows the
particle acceleration has to occur in situ, although why and how it
occurs is not understood but could involve internal shocks, magnetic
energy dissipation or a shock with the ISM.)  Self-Compton or inverse
Compton on CMB photons are also unlikely to lead to detectable levels
of emission for GLAST.  In both cases an inverse Compton luminosity
greater or equal to the synchrotron luminosity requires a magnetic
field $<3~\mu$G, a hundred times below equipartition.  Therefore,
although these observations undoubtedly show the presence of particles
of high-energies in microquasar jets, the gamma-rays emitted in those
conditions are not likely to be detected by GLAST.

\subsection{Large scale jet-ISM interaction} The dissipation of the
jet power in the ISM is a possible source of gamma-ray emission for
GLAST.  The energy involved can be substantial. For example, radio
observations show the a.u.-scale compact jet of Cyg X-1 is prolonged
into a parsec-scale structure associated with the termination shock.
The inferred power inconspicuously transported to the large scales is
comparable to the bolometric luminosity of the binary
\cite{2005Natur.436..819G}. Jets can therefore quietly inject large
amounts of high energy particles into the ISM, which might be traced
by their gamma-ray emission. Heinz \& Sunyaev
\cite{2002A&A...390..751H} estimate their contribution could reach
10\% of the Galactic cosmic ray luminosity.  If there is indeed a
significant component of high energy nucleons, a jet interacting with
a nearby molecular cloud (effectively modeled as an accelerator +
beam dump in \cite{2005A&A...432..609B}) would create pions.  The
subsequent decay, bremsstrahlung and inverse Compton emission can be
detected by GLAST, depending upon the jet power and composition, but
also upon the duty cycle of the ejection process, the distance of the
cloud to the source etc. As an example, the emission predicted in
Fig.~12 of \cite{2005A&A...432..609B} is detectable within a year for
a microquasar at a distance of 1~kpc (e.g. A0620-00, XTE J1118+480,
the latter having the advantage of a large Galactic latitude
$b$=+62.3$^{\rm o}$). Such observations would give new clues as to the
content and power of relativistic jets in binaries.

\subsection{Gamma-ray spectral states and major ejections} Gamma-ray
emission originating closer in to the compact object can also be
expected. There is reasonable evidence from CGRO observations for soft
power-law tails (spectral slope $\alpha\approx$-1.5) extending beyond
100~keV, up to several MeV in some X-ray binaries
\cite{1998ApJ...500..899G}. These soft tails are a defining property
of the very high state (or steep power law state) together with
significant thermal emission (around a keV) and fast variability
(QPOs) \cite{2006ARA&A..44...49R}. This is most clearly seen in the
high state of Cyg X-1 where the power-law extends to 10~MeV.
Extrapolating shows the 100~MeV emission was beyond the reach of EGRET
but should be detectable by GLAST within days. Cyg X-1 spends 90\% of
its time in the hard state where there is a hint for a similar, but
fainter, power-law component that could be detected in a year by
GLAST. These power-laws can be produced in plasmas where a fraction of
the accretion energy goes into non-thermal channels
\cite{2004PThPS.155...99Z}.  Models predict a cutoff in the GLAST
energy range either because there is a maximum electron energy or
because pair production sets in when the plasma compacity is high.

Changes in X-ray spectral states can therefore be surmised to be
associated with changes in HE gamma-ray luminosity, the very high
X-ray state (resp. hard X-ray state) involving high (resp. low) levels
of gamma-ray emission. Interestingly, spectral state changes from hard
to very high X-ray states have been conjectured to be associated with
major relativistic ejections \cite{2004MNRAS.355.1105F} and one should
note, perhaps naively, that if jets are composed of $e^+e^-$ pairs
then there has to be some gamma-ray emission linked to the pair
production, if only at a few MeV. Radio/IR observations of discrete
ejections in GRS 1915+105 show non-thermal emission cooling with
expansion. Extrapolating back to early times, the fluence expected in
a day by Atoyan \& Aharonian \cite{1999MNRAS.302..253A} in HE gamma-rays is detectable by
GLAST. Gamma-ray monitoring of outbursting binaries or sources such as
GRS 1915+105, a task that is well-suited to GLAST, can therefore shed
light on how spectral state changes relate to ejection events.

\section{The observational status: gamma-ray binaries}

\subsection{The view from space} Although the above (should)
demonstrate that there are reasonable grounds to expect HE gamma-rays
from compact binaries, observational confirmation has proved elusive
and when found, arguably disconcerting. Several tentative associations
of binaries with EGRET sources were made based on positional
coincidence and/or the detection of variability, with the source 2CG
135+01 figuring prominently as the first and most secure: follow-up
observations carried out after the initial COS B discovery had
revealed a high-mass X-ray binary, LS I+61 303, in a 26 day elliptical
orbit showing periodic radio outbursts \cite{1978Natur.272..704G}. The
latter feature being rare, this highlighted the system as a plausible
counterpart.  Yet, although variability was reported in the EGRET
data, neither this nor the position were enough to formally identify
the two (the HE variation not being tied to variability at other
wavelengths and the stellar counterpart being localized only in
between the 95\% and 99\% confidence contours of the HE source). The
limited angular resolution combined with the strong underlying
Galactic diffuse HE emission resulted in error boxes of a several tens
of arcmins, too large to pick up the needle in the haystack of
possible Galactic Plane counterparts.

\subsection{The view from the ground}Breakthrough observations were
obtained by the ground-based Cherenkov telescopes operated by the HESS
and MAGIC collaborations. These observe at a higher threshold ($\geq
100$~GeV) but benefit from a larger collecting area and a better
angular resolution. Three binaries were detected: PSR B1259-63, LS
5039 and LS I+61 303
\cite{2005A&A...442....1A,2005Sci...309..746A,2006Sci...312.1771A}.
The latter two had tentative EGRET associations but not PSR B1259-63,
possibly because the HE gamma-ray emission is highly variable along
the 3.5 year orbit and confined to a short period around periastron
passage. The next passage occurs this year, too early for GLAST so
searches may have to wait 2010 (unfortunately, periastron passage
cannot be observed by HESS before 2014). The localizations are much
more precise: for instance, LS 5039 is coincident with HESS J1826-148
within the positional uncertainty of 30'', excluding a nearby SNR and
a pulsar that were within the error box of the EGRET source 3EG
1824-1514. More importantly, all three binaries display variability.
In LS 5039, it was demonstrated that the $>$100~GeV flux is strongly
modulated on the orbital period \cite{2006A&A...460..743A} and there
is little doubt that the fluxes measured in PSR B1259-63 and LS I+61
303 also depend on orbital phase.

\subsection{Gamma-ray binaries}All three binaries have high mass O or
Be type companions with compact objects in eccentric orbits.  The
X-ray output from these binaries is about 10$^{34}$~erg/s, rather low
in itself compared to typical X-ray binaries, and smaller or
comparable to the emission above 100~MeV. These systems are therefore
gamma-ray loud, a first surprise. All of them display radio emission,
resolved in LS 5039 and LS I+61 303 as collimated outflows on
milliarcsecond scales, immediately suggesting a microquasar nature
\cite{2000Sci...288.2340P}.  X-ray and radio variability, when (and
if) present is of limited amplitude (odd in accreting binaries) and
occurs on the orbital timescale.  The overall spectral energy
distributions are similar for all three systems, showing (in $\nu
F_\nu$) a rising spectrum from radio to X-rays flattening around an
MeV and reaching energies of several TeV.  However, PSR B1259-63 is
not a microquasar but a young 48 ms pulsar with a spin-down power
$\approx 10^{36}$~erg~$s^{-1}$.  The relativistic pulsar wind is
sufficient to quench any wind accretion and the emission is thought to
arise from particles accelerated where the pulsar and stellar wind
interact \cite{1997ApJ...477..439T}. This paints a rather different
picture than accretion-powered scenarios.

\subsection{Gamma-ray binaries as compact pulsar wind nebulae} 
Why the observational properties of the three gamma-ray binaries
should bear any resemblance is disconcerting unless all are actually
rotation-powered by a young pulsar \cite{2006A&A...456..801D}. This
can explain the low, steady level of emission, conceivably modulated
as the pulsar moves around its orbit. Pulsed emission would be
absorbed in the stellar wind because of the smaller orbital
separations in LS 5039 and LS I+61 303 (as observed near periastron in
PSR B1259-63). The pulsar wind is confined by the stellar wind to a
cometary nebula pointing away from the massive companion, producing
the collimated radio outflow. Radio VLBI observations of LS I+61 303
recently reported by \cite{2006smqw.confE..52D} are consistent with
this picture, showing a periodic sweep of the radio tail with orbital
phase which appears irreconcilable with an accretion-powered jet: LS
I+61 303 is almost certainly powered by a pulsar. The small-scale
radio morphology of LS 5039 has been successfully modeled as a pulsar
wind nebula \cite{2006A&A...456..801D} but this does not provide the
same level of certainty in the absence of observations at other
orbital phases.  Other models have proposed the emission arises in a
relativistic jet powered by accretion
\cite{2006ApJ...643.1081D,2006A&A...451..259P}, but that hypothesis
seems rather uneconomical to this author. At this stage, it seems more
probable that HE gamma-ray emission from accretion-powered binaries
has yet to be detected and that, when this will be achieved, their
observational properties will be clearly different from those of the
rotation-powered binaries.

\section{GLAST studies of rotation-powered binaries}

\subsection{Gamma-ray orbital modulation}That all three gamma-ray
binaries have high-mass stellar companions may be instrumental to
generate the HE emission, as these will provide copious amounts of
seed photons for inverse Compton scattering. On the other hand, the
large photon densities at UV energies also imply pair production with
TeV photons can be important. The starlight both provides a source and
a sink for gamma-rays. Gamma-ray absorption has a strong orbital
dependence as the cross-section for pair production depends on the
angle between the two photons. The effect can be dramatic on
gamma-rays emitted towards the observer and crossing head-on the path
of stellar photons. In LS 5039, the tight 4-day orbit brings the
compact object to within a stellar radius from the O6V star.  Gamma-rays
of $>$30~GeV emitted close to the compact object are modulated with
peak attenuation at superior conjunction (when the compact object is
behind the star as seen by the observer) \cite{2006A&A...451....9D}.
Such an orbital modulation has been observed in LS 5039 by HESS with a
peak and trough at the predicted orbital phases
\cite{2006A&A...460..743A}.

However, the flux at superior conjunction is not completely absorbed.
Furthermore, the $>$100~GeV spectrum varies from a soft power law at
superior conjunction (low flux) to a hard power law at inferior
conjunction (high flux), whereas pure absorption predicts a dip in the
spectrum around a TeV. Other effects must play a role. One is that a
pair cascade is initiated when the newly created $e^+e^-$ up-scatter
star photons back into the absorption range. The magnetic field has to
be low enough ($\leq$10~G) to prevent synchrotron losses from
dominating. The radiated energy is redistributed below the pair
production threshold ($<$30~GeV) i.e. in the GLAST range.
Calculations show an anti-correlation of the HESS and GLAST
light-curves \cite{2007A&A...464..259B}, detectable within a year
\cite{2006smqw.confE..68D}, that would prove the existence of a
cascade.

A second effect is that inverse Compton scattering is also
anisotropic. For example, this will decrease the gamma-ray flux
around inferior conjunction, when the (incoming) star photons and
(outgoing) gamma-ray photons both go towards the observer: in this
configuration the energy of the outgoing photon and the cross-section
for IC are small. There actually is a hint of a dip in the HESS
light-curve at this phase.

\subsection{Probing pulsar winds}
Besides these geometrical effects, the efficiency with which
gamma-rays are emitted may also change along the orbit. The variations
in the $>$100~GeV flux observed by HESS and MAGIC in LS I+61 303 and
PSR B1259-63 do not match the expected light-curve for pure absorption
(whose effect is marginal due to the wider orbits) and so must be
intrinsic to the emission process \cite{2006astro.ph..8262D}. LS I+61
303 may prove a Rosetta stone for this problem as its orbit is both
wide enough (0.2-0.7 a.u.)  to avoid most cascading and absorption,
but short enough (26 days) to allow for detailed studies over many
orbits.  MAGIC reports a minimum in flux close to periastron (which is
close to inferior conjunction, where absorption is minimal) and a
maximum towards apastron. The explanation is straightforward with a
compact pulsar wind nebula. The pulsar wind is contained by the
stellar wind of the Be companion. At periastron, the ram pressure of
the dense equatorial wind crushes the PWN nebula to a small distance
from the pulsar. The magnetic field at the shock location is strong
and particles lose energy rapidly to synchrotron emission without
radiating much inverse Compton above $>$10~GeV. At apastron, the
stellar wind is polar and diffuse, implying a large shock distance and
weaker magnetic field, allowing for higher particle energies emitting
more HE inverse Compton gamma-rays.  Hence, phase-resolved spectral
energy distributions from X-rays to VHE gamma-rays can yield
information on the magnetic field at different locations, forming a
probe of the relativistic wind as a function of distance to the
pulsar.

\subsection{Population studies of gamma-ray binaries}
The pulsar spin-down timescale is short ($\sim 10^5$ years for PSR
B1259-63) and can only power the binary emission for a brief period of
time.  Accretion from the stellar wind is then no longer quenched by a
pulsar wind and an X-ray pulsar turns on. Hence, gamma-ray binaries
are the progenitors of the longer-lived, accretion-powered high-mass
X-ray binaries (HMXBs).  Because most of their output is in
gamma-rays, GLAST all-sky observations provide a unique way of
identifying these progenitors and studying them as a population,
constraining the birth rate and evolution of HMXBs.  Population
synthesis calculations find there should be around 30 active gamma-ray
binaries in order to match the present-day population of HMXBs,
assuming the rotation-powered phase lasts only 10$^4$ years
\cite{1989A&A...226...88M}.  They should be visible throughout the
Galaxy if their luminosity in the GLAST band is comparable to that of
the three known systems ($\sim 10^{35}$ erg~s$^{-1}$). This estimate
does not take into account the reduction in sensitivity to be expected
in the Galactic Plane due to the diffuse emission. In this respect,
the Magellanic Clouds, which harbor a comparatively very large
population of HMXBs, might also make for good dedicated studies
despite being more distant.






\begin{theacknowledgments}
  I thank the organizers of the 1st GLAST symposium for their
  invitation and acknowledge support from the {\em Agence Nationale de la Recherche}.
\end{theacknowledgments}



\bibliographystyle{aipprocl} 

\bibliography{dubusglast_bib}

\begin{thebibliography}{10}
\providecommand{\enquote}[1]{``#1''}
\expandafter\ifx\csname url\endcsname\relax
  \def\url#1{\texttt{#1}}\fi
\expandafter\ifx\csname urlprefix\endcsname\relax\def\urlprefix{URL }\fi

\bibitem{2002Sci...298..196C}
S.~{Corbel et al.}, \emph{Science} \textbf{298}, 196--199 (2002).

\bibitem{2005Natur.436..819G}
E.~{Gallo}, R.~{Fender}, C.~{Kaiser}, D.~{Russell}, R.~{Morganti},
  T.~{Oosterloo}, and S.~{Heinz}, \emph{\nat} \textbf{436}, 819--821 (2005).

\bibitem{2002A&A...390..751H}
S.~{Heinz}, and R.~{Sunyaev}, \emph{\aap} \textbf{390}, 751--766 (2002).

\bibitem{2005A&A...432..609B}
V.~{Bosch-Ramon}, F.~A. {Aharonian}, and J.~M. {Paredes}, \emph{\aap}
  \textbf{432}, 609--618 (2005).

\bibitem{1998ApJ...500..899G}
J.~E. {Grove et al.}, \emph{\apj} \textbf{500}, 899 (1998).

\bibitem{2006ARA&A..44...49R}
R.~A. {Remillard}, and J.~E. {McClintock}, \emph{\araa} \textbf{44}, 49--92
  (2006).

\bibitem{2004PThPS.155...99Z}
A.~A. {Zdziarski}, and M.~{Gierli{\'n}ski}, \emph{Progress of Theoretical
  Physics Supplement} \textbf{155}, 99--119 (2004).

\bibitem{2004MNRAS.355.1105F}
R.~P. {Fender}, T.~M. {Belloni}, and E.~{Gallo}, \emph{\mnras} \textbf{355},
  1105--1118 (2004).

\bibitem{1999MNRAS.302..253A}
A.~M. {Atoyan}, and F.~A. {Aharonian}, \emph{\mnras} \textbf{302}, 253--276
  (1999).

\bibitem{1978Natur.272..704G}
P.~C. {Gregory}, and A.~R. {Taylor}, \emph{\nat} \textbf{272}, 704--706 (1978).

\bibitem{2005A&A...442....1A}
F.~{Aharonian et al. (HESS collaboration)}, \emph{\aap} \textbf{442}, 1--10
  (2005).

\bibitem{2005Sci...309..746A}
F.~{Aharonian et al. (HESS collaboration)}, \emph{Science} \textbf{309},
  746--749 (2005).

\bibitem{2006Sci...312.1771A}
J.~{Albert et al. (MAGIC collaboration)}, \emph{Science} \textbf{312},
  1771--1773 (2006).

\bibitem{2006A&A...460..743A}
F.~{Aharonian et al. (HESS collaboration)}, \emph{\aap} \textbf{460}, 743--749
  (2006).

\bibitem{2000Sci...288.2340P}
J.~M. {Paredes}, J.~{Mart{\'{\i}}}, M.~{Rib{\'o}}, and M.~{Massi},
  \emph{Science} \textbf{288}, 2340--2342 (2000).

\bibitem{1997ApJ...477..439T}
M.~{Tavani}, and J.~{Arons}, \emph{\apj} \textbf{477}, 439 (1997).

\bibitem{2006A&A...456..801D}
G.~{Dubus}, \emph{\aap} \textbf{456}, 801--817 (2006).

\bibitem{2006smqw.confE..52D}
V.~{Dhawan et al., in {\em Proc. of the VI Microquasar Workshop}, Sep. 18-22,
  2006, Como, Italy}, \emph{PoS (MQW6) 52}  (2006).

\bibitem{2006ApJ...643.1081D}
C.~D. {Dermer}, and M.~{B{\"o}ttcher}, \emph{\apj} \textbf{643}, 1081--1097
  (2006).

\bibitem{2006A&A...451..259P}
J.~M. {Paredes}, V.~{Bosch-Ramon}, and G.~E. {Romero}, \emph{\aap}
  \textbf{451}, 259--266 (2006).

\bibitem{2006A&A...451....9D}
G.~{Dubus}, \emph{\aap} \textbf{451}, 9--18 (2006).

\bibitem{2007A&A...464..259B}
W.~{Bednarek}, \emph{\aap} \textbf{464}, 259--262 (2007).

\bibitem{2006smqw.confE..68D}
R.~{Dubois, in {\em Proc. of the VI Microquasar Workshop}, Sep. 18-22, 2006,
  Como, Italy}, \emph{PoS (MQW6) 68}  (2006).

\bibitem{2006astro.ph..8262D}
G.~{Dubus, in {\em Proc. Vulcano workshop Frontier Objects in Astrophysics and
  Particle Physics}, May 22-27, 2006, F. Giovannelli \& G. Mannocchi (eds.),
  Italian Physical Society, Editrice Compositori, Bologna, Italy},
  \emph{(astro-ph/0608262)}  (2006).

\bibitem{1989A&A...226...88M}
E.~J.~A. {Meurs}, and E.~P.~J. {van den Heuvel}, \emph{\aap} \textbf{226},
  88--107 (1989).

\end{thebibliography}


\end{document}